\documentstyle[preprint,aps,epsf]{revtex}
 
\newcommand{\rr}{{\bf r}} 
\newcommand{\p}{{\bf p}} 
\newcommand\be{\begin{equation}} 
\newcommand\ee{\end{equation}} 
\newcommand {\hsh  } {\hspace{5mm}         }

\tightenlines
 
\begin{document} 
%\draft

\title{Folding, Design and Determination of Interaction Potentials  
Using Off-Lattice Dynamics of Model Heteropolymers} 

\author{\em Cecilia Clementi $^1$, Amos Maritan $^1$, and Jayanth R.
Banavar $^2$}
\address{}
\address{$^1$ International School for Advanced Studies (SISSA)}
\address{and Istituto Nazionale di Fisica della Materia,}
\address{Via Beirut 2-4, 34014 Trieste, Italy}
\address{}
\address{$^2$ Departement of Physics and Center for Material Physics}
\address{104 Davey Laboratory, The Pennsylvania State University,}
\address{University Park, PA 16802-USA}
\address{}

\date{\today} 
\maketitle 

\begin{abstract}
{\noindent We present the results of a self-consistent, 
unified molecular dynamics
study of simple model heteropolymers in the continuum
with emphasis on folding, sequence design
and the determination of the interaction parameters of the 
effective potential between the amino acids from the knowledge
of the native states of the designed sequences.}  
\end{abstract}

\centerline{ PACS numbers: 87.15.By, 87.10.+e }
\newpage
A fundamental problem in molecular biology is that of protein folding.
In a coarse-grained description, a protein may be thought of
as a heteropolymer made up of amino acids, twenty kinds of which 
occur in nature.  Broadly speaking, there are several 
issues that are of vital importance.  One of  these
is the direct folding problem: given the effective interactions 
between amino acids and an amino acid sequence, what is its
native state conformation or ground state structure?  This is
a crucial question because the functionality of a protein
is controlled by its native state structure.  A second issue is that
of inverse-folding or the design problem: again given the effective
interactions, what is the sequence of amino acids that would
have as its native state conformation a desired target structure?
A further requirement for a functionally useful protein is that the native 
state conformation be thermodynamically  stable and kinetically
accessible from a typical random conformation. Recent studies have
shown the role played by a folding funnel in the energy landscape
in facilitating rapid folding \cite{funnel}.  An underlying question relevant
to both the direct and inverse folding problems is how one may
determine the effective interactions between the amino acids from 
knowledge of the native state structures of several sequences,
e.g. from the Protein Data Bank (PDB).  Considerable progress
has been made in addressing each of these issues computationally
but not in an unified manner and usually within the framework
of simplified lattice models  \cite{funnel,tot}.

This letter presents a summary of the results of a comprehensive
and unified study of all of the above issues with an off-lattice model 
of heteropolymers.  Our study provides an important test of the
feasibility of the implementation of various strategies  in a
realistic, albeit simplified framework.  

A conformation  of a  chain made up of $N$ residues is 
defined by the coordinates 
$\rr_1,\ldots,\rr_N$ of beads in  
three-dimensional space.  For a real protein, the beads may, for example, 
represent the $C_{\alpha}$ atoms of the amino acids.
We consider only effective two body forces between amino acids 
obtained on integrating out the degrees of freedom associated with
the internal coordinates of each residue and the solvent.
A simple choice for the interaction potential is:
 
\be 
\label{potential} 
V_{ij}=\delta_{i,j+1}f(r_{i,j})+ \eta( 
(\frac{\sigma}{r_{ij}})^{12} - 
(\frac{\sigma}{r_{ij}})^6) , 
\ee 
 
where $r_{ij}=|\rr_i-\rr_j|$ is the inter-residue distance. 
 
The parameter $\eta$ entering this equation controls  the energy scale, 
whereas  $\sigma$ determines the interaction length between monomers. 
The values of $\sigma$ and $\eta$ have to be 
adjusted to fit both the complex interactions between the various 
groups 
of amino acids and the interactions with the solvent. 
Furthermore, these parameters could depend on the different 
types of aminoacids involved in the interaction. 

The energy function of the peptide bond is chosen to be
\be 
f(x)=a (x-d_0)^2+b (x-d_0)^4, 
\ee 
with $a$ and $b$ taken to be $1$ and $100$ respectively.
We add a quartic term to the usual quadratic one \cite{thiru}
because a plain harmonic potential could
induce energy localization in some specific
modes, significantly increasing the time needed for equilibration.

The parameter $d_0$, which represents the equilibrium distance of the 
nearest neighbors along the chain is set equal to 
$3.8\AA$, the experimental value for the mean
distance between nearest neighbor $C_{\alpha}$ atoms along the
chain in
real proteins, as determined from the PDB.

The Hamiltonian is given by: 
 
\be 
H=\sum_{i=1}^{N} \frac{\p_{i}^2}{2} + \sum_{i=1}^{N} \sum_{j>i} 
V_{i,j}. 
\ee 
 
The first term is the classical kinetic energy of the chain, where 
the 
$\p_{i}$'s are the canonical variables conjugated to the 
$\rr_{i}$'s. 

We have used Molecular Dynamics (MD) 
(entailing the integration of Newton's
laws of motion on a computer) for simulating the kinetics of the chains.
We employed an efficient and precise symplectic algorithm 
\cite{LapoAlgo}, in which one varies the
energy density $\epsilon=E/N$, which is related to the temperature \cite{PLV}.

Our computations were carried out in three stages:

1) On a lattice, one usually selects a compact conformation
and attempts to design a sequence that has this structure 
as a thermodynamically stable ground state.  Off-lattice, there
are an infinite number of conformations almost all of which are not designable
(i.e. there is no sequence which has the conformation as its ground state).
Our first goal was to select a number of compact,
designable conformations.  We accomplished this by beginning with a homopolymer
model (just one kind of amino acid) with overall attractive interactions
between pairs of monomers.  For the homopolymer case, we fixed
the parameters $\eta = 40$ and $\sigma = 6.5\AA$. 
Such a value for
$\sigma$ ensures that, in practice, two monomers significantly interact with
each other when their distance is smaller then $9$ \AA.  Such a distance
threshold is conventionally used for the bond between amino acids
and is determined by the requirement that the average number
of $C_{\alpha} - C_{\alpha}$ contacts for each amino acid is roughly
equal to the respective numbers obtained with the all-atom
definition of contacts \cite{michele_map}.

For a three dimensional homopolymer made up of 30 identical monomers,
twenty compact conformations (the radii of gyration varied between 7.52 and 
7.59 $\AA$) with low-lying energies were obtained  performing
MD simulations in a slow-cooling mode. 
The system was equilibrated after successive cooling on lowering the temperature
each time by a factor of 0.8.  The compact 
conformations were chosen so that they had little structural overlap with each 
other.   The distance $D$ between two 3-dimensional conformations is
given by
\be
D = \sqrt{\frac{1}{N} \sum_{i=1}^{N} (\rr_{i}-\rr_{i}^{,})^2}, 
\label{eq_dist}
\ee
where one structure is translated and rotated to get a minimal D \cite{kab}.  
Two conformations were assumed to be different if the $D$ between 
them exceeded $1 \AA$ based on the experimental resolution of
protein structures \cite{rxn}.

The mean distance  between a given conformation and the 19 others
ranged between 6.67$\AA$ and 7.49$\AA$.
Strikingly, the resulting structures possess secondary motifs,
especially helices (see Figure 1). 
The appearence of secondary motifs is not a general phenomenon but is
linked to the our choice of the length parameters
$\sigma$ and  
$d_0$. 
The relation between the ratio $\sigma / d_0$ and the nature of the conformation
of low-lying energy states will be discussed elsewhere \cite{next_work}.

2) We next switched to a heteropolymer model employing four types
of amino acids, two of which were predominantly hydrophobic
and thus mutually strongly attractive (on integrating the solvent
degrees of freedom) and the other two were hydrophilic.  The
composition of the 
sequence was constrained so that there were 6 amino acids each of the 
hydrophobic types and 9 amino acids each of the third and fourth kinds.
The interaction potential had the same form as before but with ten 
parameters characterizing the Lennard-Jones interactions between the
four types of amino acids
chosen by hand (see Table \ref{table1}).

Small variations in the Lennard-Jones
length parameter (5 possible values of $\sigma$ equal to 6.25, 6.5, 
7.0, 7.5 and 8.0 $\AA$ were now permitted unlike the homopolymer 
model in which just the 6.5 $\AA$ value was employed) that were
amino acid-type independent but allowed for the stabilization of 
the target native states were allowed to take into account approximately
the diversity of sizes of amino acids in nature.
A simplified design procedure due to Shakhnovich and Gutin \cite{des.sh}
was carried out in which the designed sequences are chosen so as to
minimize the energy in the target conformation and entailed an optimal 
assignment of  amino acid type to each monomer and independently a choice
of the $\sigma$ parameter for each $i-j$ pair with $|i-j| > 1$.
The design procedure was carried out for each of the twenty 
conformations deduced using the homopolymer model and was validated
by detailed simulations which showed that the designed sequences
do indeed have the target conformations as their ground states.
We slowly cooled each designed sequence
several times (typically 50) from different random initial
conditions. From this procedure, we confirmed that the
target conformations are indeed the lowest energy structures. 
These cooling simulations also generate a set of alternative, 
higher energy, metastable conformations (2-5 for each sequence) that, 
when perturbed, ``decay'' to the global minimum conformation (the target 
structure).
The energy landscape is modified by the design procedure and a 
folding funnel, that promotes thermodynamic stability and kinetic 
accessibility is created as shown for one of the designed sequences 
in Figure 2.

3)  We then set about to determine the effective parameters of the potential 
of interaction between the four kinds of amino acids using the knowledge
of the test bank of sequences and their known native structures. The basic
idea is to require  that the
energy of the designed sequences in their ground states be less than
their energies in alternative conformations.  This is simply  a consistency
requirement for the definition of a ground state (native state).  
Recently, we have carried out studies \cite{hp.ref1} of an optimization method 
for the determination of effective potential energies of interaction
and have extensively tested 
it on lattice models.  The 
method selects the optimal parameters such that the smallest energy gap
(chosen among the set $\Omega_s$ of sequences 
$\{\sigma_s\}_{s=1,\dots,20}$ in the training set)
between the energy of the sequence in its 
native conformation $\Gamma_s^n$ and the (higher in energy)
minimum energy alternative one 
is as large as possible.
This additionally promotes thermodynamic stability and may be implemented by
defining a cost function $F_{gap}$:
\be 
\label{maxming}
F_{gap} = -min_{(\sigma_s \in \Omega_{s})} min_{(\Gamma \neq 
\Gamma_s^n)}\frac{E(\Gamma,\sigma_s) - E(\Gamma_s^n,\sigma_s) }
{|E(\Gamma_s^n,\sigma_s)|} 
\ee 
and by choosing the parameters values of the potential that minimize this
cost function.
Note that  this is a  slightly modified version of the equation 
in \cite{hp.ref1} -- the key new feature is the presence of the
denominator $|E(\Gamma_s^n,\sigma_s)|$  which serves to rescale the 
energy gap associated with a given sequence with respect to its ground 
state energy.

In lattice models
amenable to exact enumeration, all conformations other than that of the 
native state can be conveniently used as alternative or decoy conformations.
In an off-lattice model, there are potentially an infinite number of
such decoy conformations.  
We started by taking as a set of alternative trial structures, the native-like
ones obtained from the homopolymer studies and metastable structures. 
We thus used 90 different decoy 
conformations -- 19 of the 20 basic structures obtained from
the homopolymer model, excluding the native structure itself,  and 71
from the alternatives generated by the repeated cooling process (as described
above).
We proceeded to  determine rough values of the potential parameters 
that minimize  the cost function (eqn. 5) with respect to these 
set of decoy structures. 
The generic Lennard-Jones form with unspecified parameters 
$\eta$ and $\sigma$ was used as the trial potential function for the 
interactions between amino acids
with one of the $\eta$ values fixed to its true value 
in order to set the energy scale.

A key ingredient for the success of the procedure is the use of decoy 
conformations that are significant competitors to the native state in 
housing the given sequence.
To add relevant conformations to the decoy set, we used the extracted
parameter values to slowly cool each sequence about 5 times.
Initially, when  non-optimal values of parameters are used, the simulations
lead to lowest energy conformations  that differ from the true ones
for almost  all the sequences.
We used these as additional decoy structures in order to iteratively 
refine the parameters of the potential.
We iterated the procedure until it converges self-consistently,
i.e. until a cooling 
simulation with extracted parameters leads to the true ground state.
This procedure converges very nicely and yields values
in excellent accord with the chosen potential parameters (Figure 3).
The final iteration step is obtained using a decoy set of 1631
structures (i.e. 19 of the 20 basic structures and 1612 alternative
ones).

Taken together, these steps lead to a unified and entirely self-consistent
description of perhaps the simplest off-lattice model of heteropolymer
chains and opens the way for similar studies of small segments of real
proteins.  An important feature of the study is that a simple known
potential was used  for the design studies and  therefore 
facilitated the verification of the potential parameters that were
subsequently determined.  This luxury is not present for similar studies
on real proteins, for which the potential energies of interactions
between amino acids are truly unknown.

In summary, we have carried out a comprehensive study of the principal issues 
involved in the protein folding problem using a simple off-lattice model in 
three dimensions.  
Our results include the observation of secondary motifs in 
the native state structures, the creation of a folding funnel in the energy 
landscape of designed sequences and successful tests of folding, design and the 
extraction of parameters characterizing the interaction potential between the 
amino acids.  An application of these strategies to coarse-grained models of 
short proteins seems eminently feasible. 
 
One of us (CC) is grateful to Giovanni Fossati for helpful advice.
This work was supported by grants from NATO, NASA, The Center
for Academic Computing and the Applied Research Laboratory
at Penn State.

%-------------------------------------------------------------
%$$$$$$$$$$$$$$$$$$ REFERENCES $$$$$$$$$$$$$$$$$$$$$$$$$$$$$$$
%-------------------------------------------------------------

\newpage
%%%%%%%%%%%%%%%%%%%%%%%%%%%%%%%%%%%%%%%%%%%%%%%%%%%%%%%%%%%%%%%%%%%%%%%%%%
% table captions
%%%%%%%%%%%%%%%%%%%%%%%%%%%%%%%%%%%%%%%%%%%%%%%%%%%%%%%%%%%%%%%%%%%%%%%%%%

\begin{center}
TABLE CAPTIONS
\end{center}

\noindent{\footnotesize
{\bf Table \ref{table1}}  \hsh
Values of parameters $\eta$ of the Lennard Jones interaction potential used
in the heteropolymer model for the four types of aminoacids.
}

%%%%%%%%%%%%%%%%%%%%%%%%%%%%%%%%%%%%%%%%%%%%%%%%%%%%%%%%%%%%%%%%%%%%%%%%%%
% figure captions
%%%%%%%%%%%%%%%%%%%%%%%%%%%%%%%%%%%%%%%%%%%%%%%%%%%%%%%%%%%%%%%%%%%%%%%%%%

\begin{center}
FIGURE CAPTIONS
\end{center}

\noindent{\footnotesize
{\bf Fig.1} \hsh
One of the compact structures obtained for the homopolymer model.
Note that the structure exhibits a helix having a complete turn in
10--12 beads, whereas in naturally occuring proteins, helices 
have  3.6 residues per turn. The designed sequence for this conformation
is: 

3 1 1 1 1 4 4 4 2 3 4 4 3 1 2 2 1 2 2 2 4 4 3 4 3 3 3 3 3 4.

\noindent There are $65$ contacts having a $\sigma$ value for 
equilibrium distance equal to $6.25$ \AA, $41$ with $\sigma$ 
equal to $6.5$ \AA, $12$ with $7$ \AA, $15$ with $7.5$ \AA, 
and $48$ with $8$ \AA.
}
\medskip
\medskip

\noindent{\footnotesize
{\bf Fig.2} \hsh
The energy landscape of one of the designed sequences derived from
the conformations obtained during numerous dynamical runs of slow  cooling.
The energy of each conformation is plotted as a function of its distance
(see eqn. \ref{eq_dist}) from  two fixed ``reference" conformations.
}
\medskip
\medskip

\noindent{\footnotesize
{\bf Fig.3} \hsh
Extracted parameters of the potential versus the true parameters as obtained in
the last iteration. Dark circles represent the $\eta$ parameters, whereas
the light squares denote the  $\sigma$ parameters.
}

\vfill\eject

\begin{table}
%\begin{tabular}{|p{2.4cm}|p{1cm}|p{1cm}|p{1cm}|p{1cm}|}
\begin{tabular}{|c|c|c|c|c|}
\hline
 {residue type } & { 1 } & { 2 } & { 3 } & { 4 }  \\
\hline
 { 1 }  & { 40 } & { 30 } & { 20 } & { 17 }  \\
 { 2 }  & { 30 } & { 25 } & { 13 } & { 10 }  \\
 { 3 }  & { 20 } & { 13 } & { 5 } & { 2 }  \\
 { 4 }  & { 17 } & { 10 } & { 2 } & { 1 }  \\
\hline
\end{tabular}
\caption{}
\label{table1}
\end{table}


\begin{thebibliography}{99} 
 
\bibitem{funnel}
H. Frauenfelder, S.G. Sligar and P.G. Wolynes, Science {\bf 254},
1598 (1991); 
M.R. Betancourt and J.N. Onuchic, J. Chem. Phys. {\bf 103}, 
773 (1995); J.D. Bryngelson, J.N. Onuchic and P.G. Wolynes, Proteins,
Struct., Funct. and Genetics
{\bf 21}, 167 (1995); J.N. Onuchic, P.G. Wolynes and N.D. Socci, 
Proc.  Natl. Acad. Sci. U.S.A. {\bf 92}, 3626 (1995);
N.D. Socci, J.N. Onuchic, and P.G. Wolynes, J. Chem. Phys. 
{\bf 104}, 5860 (1996); J.N. Onuchic, Z. Luthey-Schulten 
and P.G. Wolynes, Ann. Rev. Phys. Chem. {\bf 48}, 545 (1997). 

\bibitem{tot} 
See, for example,
A. Sali, E. Shakhnovich and M. Karplus, Nature {\bf 369}, 248 (1994); J.
Mol. Biol. {\bf 235}, 1614 (1994);
E. Shakhnovich and A.M. Gutin, Nature {\bf 346}, 773 (1990);  Proc.
Natl. Acad. Sci. U.S.A. {\bf 90}, 7195 (1993);
E. Shakhnovich, G. Farztdinov, A.M. Gutin and M.Karplus, Phys. Rev.
Lett. {\bf 67}, 1665 (1991);
E. Shakhnovich, Phys. Rev. Lett. {\bf 72}, 3907 (1994);
V.Abkevich, A.M. Gutin and E. Shakhnovich, J. Chem. Phys. {\bf 101},
6052 (1994); J. Mol. Biol. {\bf 252}, 460 (1995);
K.A. Dill, S. Bromberg, S. Yue, K. Fiebig, K.M. Yee, P.D. Thomas and H.S.
Chan, Protein Science {\bf 4}, 561 (1995);
T. Garel, H. Orland and D. Thirumalai, New Developments in Theoretical
Studies of Proteins, R. Elber (ed.), World Scientific, Singapore (1996);
K.F. Lau and K.A. Dill, Macromolecules {\bf 22}, 3986 (1989);
G.M. Crippen, Biochemistry {\bf 30}, 4232 (1991);
V.N. Maiorov and G.M. Crippen, J. Mol. Biol. {\bf 227}, 876 (1992);
R.A. Goldstein, Z.A. Luthey-Shulten and P.G. Wolynes, Proc. Natl. Acad. Sci.
U.S.A. {\bf 89}, 9029 (1992);
P.D. Thomas and K.A. Dill, J. Mol. Biol. {\bf 257}, 457 (1996);
L.A. Mirny and E.I. Shakhnovich, J. Mol. Biol. {\bf 264}, 1164 (1996);
J. M. Deutsch and T. Kurosky, Phys. Rev. {\bf E 56}, 4553 (1997).
We extend our apologies to other authors whose papers could not be cited
for lack of space.

\bibitem{thiru}
T. Veitshans, D.K. Klimov and D. Thirumalai,
Folding \& Design (in press);
R. M. Scheek, W.F. Van Gunsteren and R. Kaptein,
Methods in Enzymology {\bf 177}, 204 (1989).

\bibitem{LapoAlgo}
L. Casetti,
Physica Scripta  {\bf 51}, 29 (1995).

\bibitem{PLV}
J. L. Lebowitz, J. K. Percus and L. Verlet,
Phys. Rev. {\bf 153}, 250 (1967).

\bibitem{michele_map}
M. Vendruscolo, E. Kussell, and E. Domany, 
Folding and Design {\bf 2}, 295 (1997);
L. Mirny, and E. Domany,
Proteins: Struct., Funct. and Gen. {\bf 26}, 391 (1996).

\bibitem{rxn}
see, for example:
Z. Dauter, L.C. Sieker, K.S. Wilson,  
Acta Cryst. {\bf B 48}, 42 (1992).

\bibitem{des.sh} 
E. Shakhnovich and A.M. Gutin, 
Protein Eng. {\bf 6}, 793 (1993);
E. Shakhnovich, 
Phys. Rev. Lett {\bf 72}, 3907 (1994).

\bibitem{kab} 
W. Kabsch, 
Acta Cryst. {\bf A 32}, 922 (1976); 
Acta Cryst. {\bf A 34}, 827 (1978).

\bibitem{next_work}
C. Clementi, A. Maritan, J.R. Banavar,
in preparation.

\bibitem{hp.ref1} 
F. Seno, A. Maritan and J.R. Banavar, Proteins: Struct., Funct. and Genetics
(in press);
J. van Mourik, C. Clementi, A. Maritan, F. Seno and J. R. Banavar, (preprint).
 
\end{thebibliography}
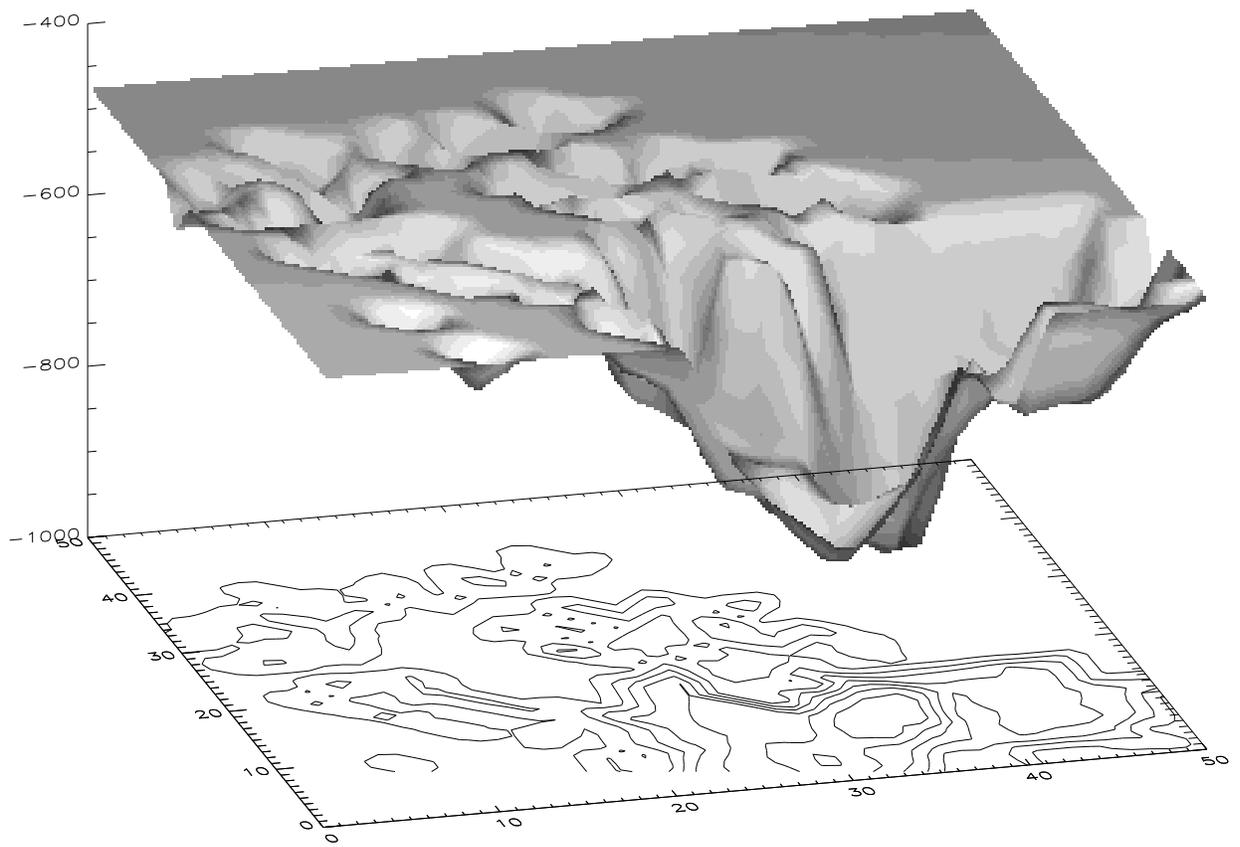
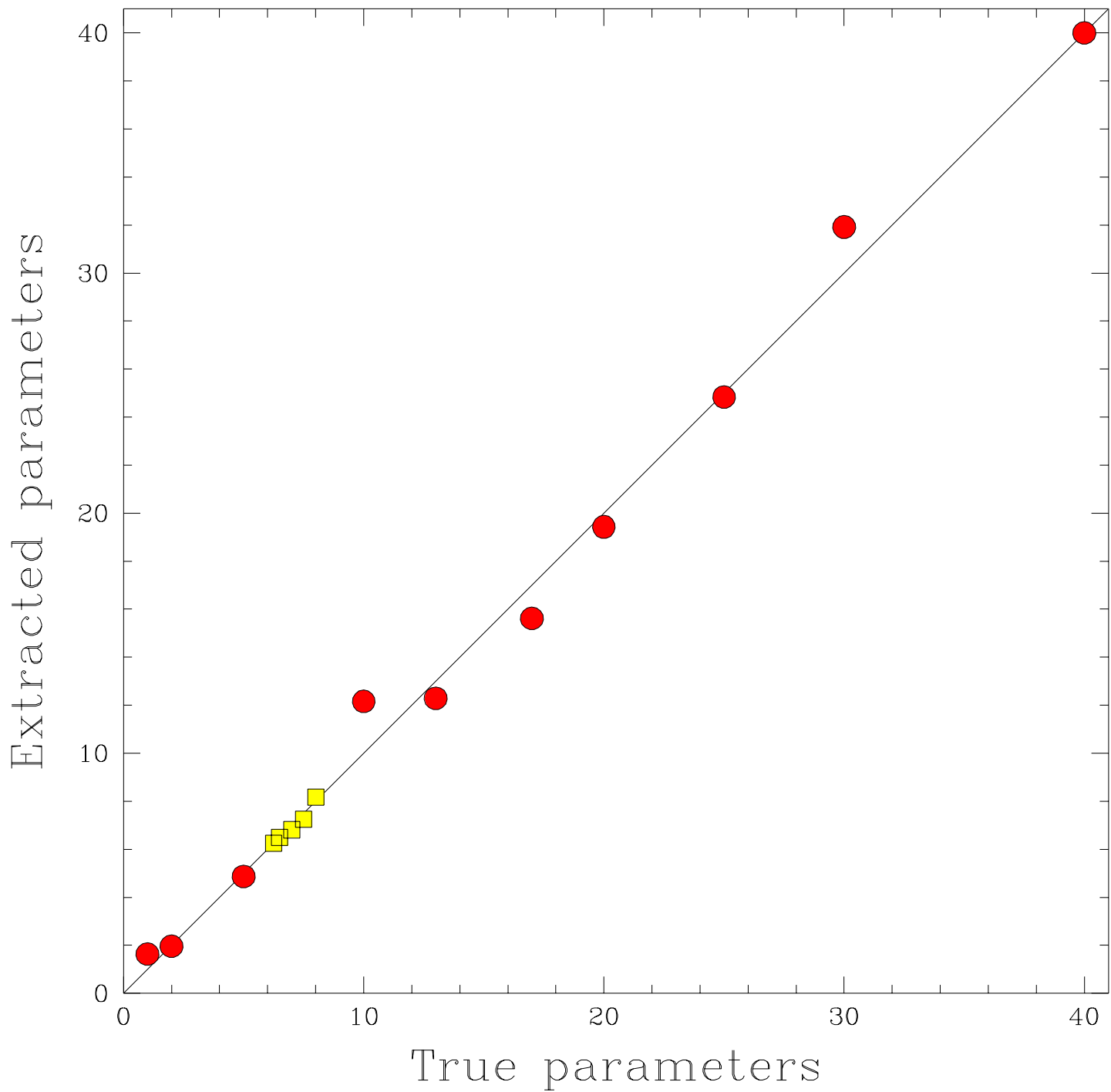
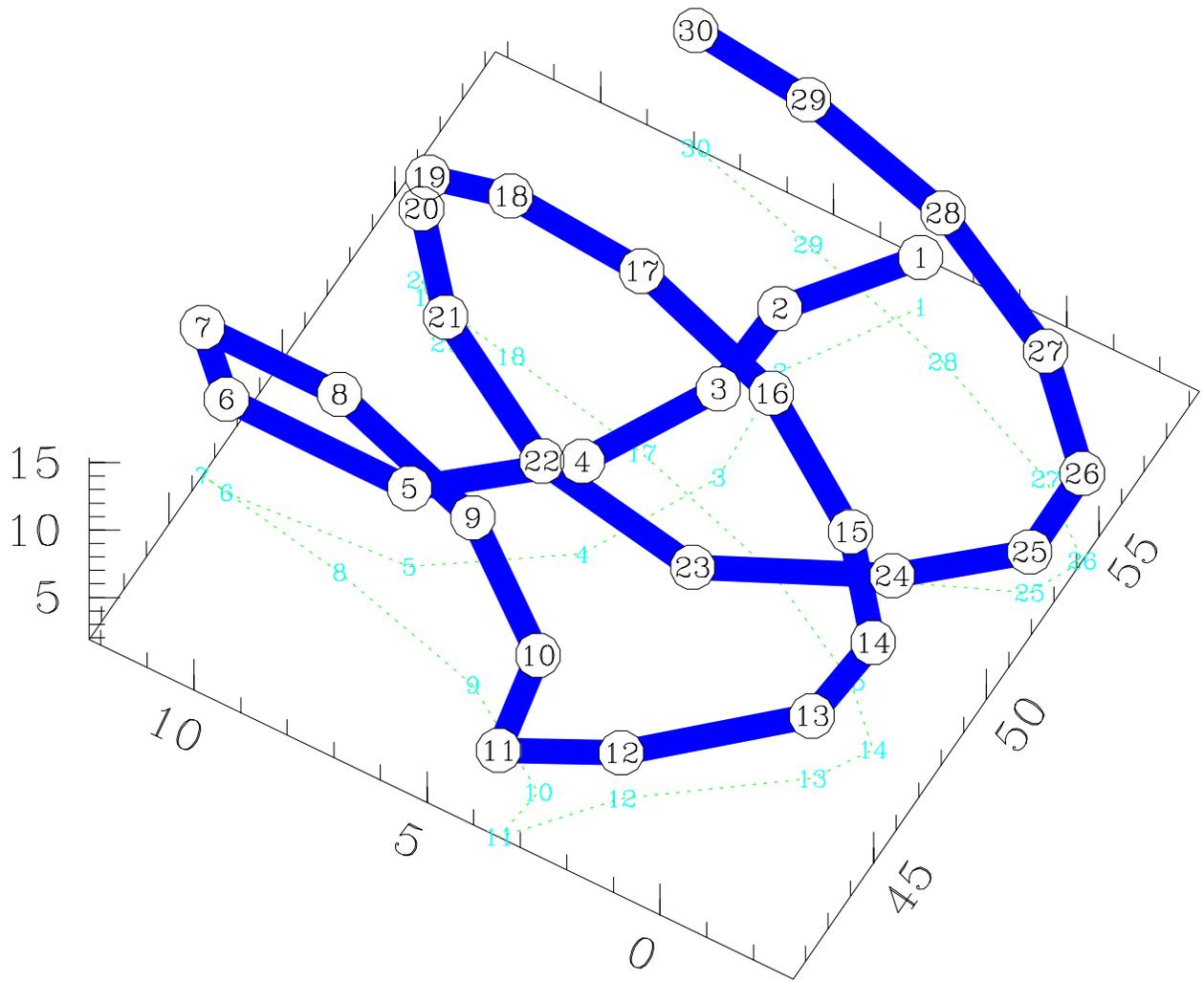
\end{document}